\documentclass[prd,aps,preprint,showpacs,nofootinbib,superscriptaddress]{revtex4}
\usepackage{epsfig}
\usepackage{amsmath}
\newcommand{\bef}{\begin{figure}[htb]\centering}
\newcommand{\eef}{\end{figure}}
\newcommand{\be}{\begin{equation}}
\newcommand{\ee}{\end{equation}}

\newcommand{\beqn}{\begin{eqnarray}}
\newcommand{\eeqn}{\end{eqnarray}}

\begin{document}


\title{Twist-three Fragmentation Function Contribution to the Single Spin Asymmetry
in $pp$ Collisions}

\date{\today}

\author{Zhong-Bo Kang}
\email{zkang@bnl.gov}
\affiliation{RIKEN BNL Research Center,
                Brookhaven National Laboratory,
                Upton, NY 11973}

\author{Feng Yuan}
\email{fyuan@lbl.gov}
\affiliation{Nuclear Science Division,
                Lawrence Berkeley National Laboratory,
                Berkeley, CA 94720}
\affiliation{RIKEN BNL Research Center,
                Brookhaven National Laboratory,
                Upton, NY 11973}

\author{Jian Zhou}
\email{jzhou@lbl.gov}
\affiliation{Nuclear Science Division,
                Lawrence Berkeley National Laboratory,
                Berkeley, CA 94720}

\begin{abstract}
We study the twist-three fragmentation function contribution to
the single transverse spin asymmetries in inclusive hadron production
in $pp$ collisions, $p^{\uparrow}p\to h+X$. In particular,
we evaluate the so-called derivative contribution which
dominates the spin asymmetry in the forward direction of
the polarized proton.
With certain parametrizations for the twist-three fragmentation
function, we estimate its contribution to the asymmetry of $\pi^0$ production at
RHIC energy. We find that the contribution
is sizable and might be responsible for the big difference between
the asymmetries
in $\eta$ and $\pi^0$ productions observed by the STAR
collaboration at RHIC.
\end{abstract}

\maketitle

Single-transverse spin asymmetries (SSAs) in hadronic processes,
such as in the single inclusive hadron production in single
transversely-polarized nucleon-nucleon scattering, $p^{\uparrow} p\to hX$,
have attracted much interests from both experimental and theoretical
sides in the last few years, and great progress has been made
in understanding the underlying physics \cite{D'Alesio:2007jt}.
Although it is a simple observable, defined as the spin asymmetry
when one flips the transverse spin of one of the hadrons involved in the scattering:
$A_N=(d\sigma(S_\perp)-d\sigma(-S_\perp))/(d\sigma(S_\perp)+d\sigma(-S_\perp))$,
it is far more complicated to explain in the fundamental theory of strong
interaction.
It usually represents a correlation between the transverse
polarization vector $S_\perp$ of one of the hadrons and the transverse
momentum $P_{h\perp}$ of the final-state hadron in the differential
cross section. For example, in the process $p^{\uparrow} p\to hX$,
it is the correlation between the polarization vector $S_\perp$ of
the incoming nucleon
and the transverse momentum $P_{h\perp}$ of the final-state hadron that generates
the asymmetry. In this paper, we will focus on the SSAs in these processes, especially for
the neutral mesons, $\pi^0$ and $\eta$ production, motivated by the
recent striking experimental observations of
large SSAs for them by the STAR
collaboration at RHIC experiments~\cite{steve, star}. In particular,
at forward direction of the polarized proton,
it was found that the SSA for $\eta$ meson is much larger than
that for $\pi^0$ meson. This also confirmed previous experimental
observations from the fixed target experiments~\cite{Adams:1997dp}. In addition to
these STAR results, both BRAHMS and PHENIX collaborations at RHIC have
observed large single transverse spin asymmetries in charged and
neutral meson production in the forward rapidity region~\cite{{:2008mi},phenix}.

In the QCD framework, there have been mainly two approaches to study
the SSAs in high energy scattering processes: the transverse momentum
dependent (TMD) parton distribution and fragmentation function approach~\cite{Siv90,Col93, Collins:1993kq,Ans94,MulTan96,BroHwaSch02,
Col02,BelJiYua02,BoeMulPij03} and
the twist-three quark-gluon correlation approach in the collinear
factorization~\cite{Efremov,qiu,Kouvaris:2006zy,{Koike:2009ge},Kang:2008qh,Eguchi:2006qz}.
In the TMD approach, it is required to have an additional hard
momentum scale besides the transverse momentum $P_{h\perp}$
of the hadron, such as $Q^2$, the momentum transfer square
of the virtual photon in semi-inclusive deep inelastic scattering (SIDIS)
or Drell-Yan lepton pair production in $pp$ collisions, to study the
transverse momentum dependence in these processes. On the other
hand, the twist-three approach is more appropriate in the processes
that all momentum scales are much larger than the nonperturbative
scale $\Lambda_{\rm QCD}$~\cite{qiu}.  It has been shown that these two
approaches are consistent with each other
for the SIDIS and Drell-Yan processes
in the intermediate transverse momentum
region  where they
both apply~\cite{Ji:2006ub,{yuan-zhou}}.

The TMD approach
has also been used to calculate the SSA in the single inclusive hadron
production in $p^\uparrow p$ collisions in a model dependent
way~\cite{Ans94,Anselmino:2005sh,Yuan:2008tv}.
However, there has been no factorization argument based on
TMD parton distribution for
this process~\cite{Collins:2007ph}.
In this paper, we follow the collinear factorization approach
to study the single spin asymmetry coming from the twist-three
fragmentation function contribution.
In this process, there is only one large momentum scale, the transverse
momentum $P_{h\perp}$ of the final-state hadron, and the SSA is naturally
suppressed by $1/P_{h\perp}$ at large $P_{h\perp}$ as a higher-twist effect,
although how large this
behavior should manifest is not clear. At low transverse momentum,
it will ultimately fail, and a modification
has to be made to account for the experimental data~\cite{star,kang}.

In the twist-three collinear factorization approach, for
the SSAs in hadron production in
\begin{equation}
p^{\uparrow}(A)+p(B) \to h+X\ ,
\label{pp}
\end{equation}
the twist-three effects come from the distribution functions of the incoming
polarized nucleon ($A$) with momentum $P_A$
or the unpolarized nucleon ($B$) with momentum $P_B$,
or the fragmentation function for
the final state hadron ($h$) with momentum $P_h$.
Therefore, schematically, we can write down the
single transverse spin dependent differential cross section in the following
way~\cite{qiu,{Kouvaris:2006zy}},
\begin{eqnarray}
d\sigma(S_\perp)&=&\epsilon_\perp^{\alpha\beta}S_{\perp\alpha}P_{h\beta}
\int [dx] [dy] [dz]\left\{\phi_{i/A}^{(3)}(x,x')
\otimes \phi_{j/B}(y)\otimes D_{h/c}(z)\otimes
{\cal H}_{ij\to c}^{(A)}(x,x',y,z) \right.\nonumber\\
&&+\phi_{i/A}(x)\otimes \phi_{j/B}^{(3)}(y,y')\otimes D_{h/c}(z)\otimes
{\cal H}_{ij\to c}^{(B)}(x,y,y',z)\nonumber\\
&&\left.+\phi_{i/A}(x)\otimes \phi_{j/B}(y)\otimes D_{h/c}^{(3)}(z,z')\otimes
{\cal H}_{ij\to c}^{(c)}(x,y,z,z')\right\}\ ,
\label{collinear}
\end{eqnarray}
where $\epsilon^{\alpha\beta}_\perp$ is defined as
$\epsilon^{\alpha\beta}_\perp=\epsilon^{\mu\nu\alpha\beta}P_{A\mu}P_{B\nu}/P_A\cdot P_B$
with convention of $\epsilon^{0123}=1$. Here we use $\perp$
to label the transverse direction in the center of mass frame of $P_A$ and $P_B$.
In the above equation, $\otimes$ stands for the convolution in the longitudinal momentum
fractions $x$, $y$, and $z$. The leading twist parton distributions are labeled by
$\phi_{i/A}(x)$ and $\phi_{j/B}(y)$. Because hadron $A$ is transversely polarized,
we immediately see that $\phi_{i/A}$ represents the leading-twist quark transversity
distributions. However, for unpolarized hadron $B$, $\phi_{j/B}$ represents both leading
twist quark and gluon distributions. Leading twist parton fragmentation function
is represented by $D_{h/c}(z)$ where parton $c$ can be a quark or gluon.
In the above equation, the superscript ${(3)}$ represents the twist-three correlations
for the distribution functions or the fragmentation functions. Since these
twist-three functions normally involve two variables in longitudinal momentum
fractions, we have made them explicit in the above formula.
For a complete analysis, the above three terms have to be taken
into account.
The first term in the above equation, the contribution from
the twist-three parton distributions $\phi_{i/A}^{(3)}(x,x')$
from the polarized nucleon have been calculated
in Refs.~\cite{qiu,Kouvaris:2006zy,{Koike:2009ge}}.
The second term from the twist-three effect
in the unpolarized nucleon $\phi_{j/B}^{(3)}(x,x')$
is found very small in the forward region of
the polarized nucleon where the largest asymmetry has been found
experimentally~\cite{Kanazawa:2000hz}.
The third term is least known in the literature, which we will
focus in this paper. There have been attempts to formulate this part
of contribution~\cite{Koike:2002ti}.
However, a universality argument for the Collins fragmentation
function~\cite{Metz:2002iz,{Meissner:2008yf}} would indicate that the contribution
calculated in Ref.~\cite{Koike:2002ti} vanishes. This universality
has also been extended to $pp$ collisions~\cite{collins-s}.
Large azimuthal asymmetries coming from this
Collins effect have been observed in semi-inclusive hadron
production in deep inelastic scattering and di-hadron production
in $e^+e^-$ annihilation processes by the HERMES~\cite{Airapetian:2004tw}
and COMPASS~\cite{Schill:2009di}, and BELLE collaborations~\cite{Abe:2005zx}, respectively.
In a recent publication, two of us have re-examined the
twist-three fragmentation contribution to the single
spin asymmetry~\cite{yuan-zhou}. In particular, we have identified
the twist-three fragmentation function corresponding to
the TMD Collins fragmentation
function, and shown that the TMD and collinear
factorization approaches are consistent in the intermediate
transverse momentum region in the SIDIS process.
In this paper, we will extend this formalism to the twist-three
fragmentation function contribution to the above process of (\ref{pp}), the
single spin asymmetries in inclusive hadron production
in $p^\uparrow p$ scattering.

The twist-three fragmentation function which
corresponds to the universal Collins TMD fragmentation function, is defined as
\footnote{In the definition of $\hat H(z)$ in Ref.~\cite{yuan-zhou},
a factor of $1/2$ missed.}
\begin{eqnarray}
\hat H(z)&=&\frac{z^2}{2}\int\frac{d\xi^-}{2\pi}e^{ik^+\xi^-}\frac{1}{2}\left\{
{\rm Tr}\sigma^{\alpha +}\langle 0|\left[iD_T^\alpha+\int_{\xi^-}^{+\infty} d\zeta^- gF^{\alpha+}(\zeta^-)\right]
\psi(\xi)|P_hX\rangle\right.\nonumber\\
&&\times \left.\langle P_hX|\bar\psi(0)|0\rangle+h.c.\right\} \ ,
\label{hzdef}
\end{eqnarray}
where $k^+=P_h^+/z$. This function can also be written as
the transverse momentum moment of the Collins fragmentation
function $H_1^\perp(z, p_T^2)$,
\begin{equation}
\hat{H}(z)=\int d^2p_T\frac{|\vec{p}_T|^2}{2M_h}H_1^\perp(z, p_T^2)\, .
\label{hz}
\end{equation}
In the above definition Eq.~(\ref{hzdef}), we have set the hadron momentum
in the $\hat z$ direction. The index $\alpha$ is in the transverse direction
perpendicular to the hadron momentum direction, and is not
considered to be summed up. In order to
uniquely define this transverse component (index), we introduce
a light-like vector $n_h^\mu\propto (P_h^0,-\vec{P}_h)$ with
normalization that $n_h\cdot P_h=1$\footnote{In the low
transverse momentum semi-inclusive hadron
production in DIS process, $n_h$ will be proportional to
the incoming nucleon momentum as we used in Ref.~\cite{yuan-zhou}.
In current study, there is no natural available momentum
for this choice. We can also perform the calculations without
choosing the vector $n_h$, but keeping all transverse component
in Eq.~(\ref{hzdef}) perpendicular to the hadron momentum. This will lead
to the same results.}.
Here and in the following, we use subscript ``$T$" to distinguish the transverse
direction perpendicular to $P_h$ from ``$\perp$" in Eq.~(2) for transverse
direction perpendicular to $P_A$ and $P_B$.
Therefore, the  transverse component for any vector $p^\mu$ can be defined
as $p_T^\mu=p^\mu-p\cdot n_h P_h^\mu-p\cdot P_h n_h^\mu$.
From this definition, we can immediately see that $p_T$ is
space-like, and perpendicular to $\vec{P}_h$:
$\vec{p}_T\cdot \vec{P}_h=0$.

As discussed in Ref.~\cite{yuan-zhou}, the above twist-three
fragmentation function belongs to more general two-variable dependent
twist-three fragmentation function defined as,
\begin{eqnarray}
\hat H_D(z_1,z_2)&=&\frac{z_1z_2}{2}\int\frac{d\xi^-d\zeta^-}{(2\pi)^2}e^{ik_2^+\xi^-}e^{ik_g^+\zeta^-}
\frac{1}{2}\left\{{\rm Tr }\sigma^{\alpha +}\langle 0|iD_T^\alpha(\zeta^-)\psi(\xi^-)
|P_hX\rangle\right.\nonumber\\
&&\times \left.\langle P_hX|\bar\psi(0)|0\rangle+h.c.\right\} \ ,
\label{hd}
\end{eqnarray}
where $k_i^+=P_h^+/z_i$ and $k_g^+=k_1^+-k_2^+$.
Similarly, we can define a $F$-type fragmentation
function $\hat{H}_F(z_1,z_2)$ by replacing $D_T^\alpha$
with $F^{+\alpha}$ in Eq.~(\ref{hd}). $D$-type and $F$-type
functions are related to each other,
\begin{eqnarray}
\hat H_D(z_1,z_2)&=&PV\left(\frac{1}{\frac{1}{z_1}-\frac{1}{z_2}}\right) \hat H_F(z_1,z_2)+
\delta\left(\frac{1}{z_1}-\frac{1}{z_2}\right) \frac{z_2}{z_1}\hat H(z_1) \ .
\end{eqnarray}
From this equation, we can see that $\hat H_D$ is more singular
than $\hat H_F$ at $z_1=z_2$. Moreover, recent study has shown that $\hat H_F(z_1,z_2)$
vanishes when $z_1=z_2$~\cite{Meissner:2008yf}. Therefore, it is more convenient to express
the cross section in terms of $\hat H(z)$ and $\hat H_F(z_1, z_2)$, which we will follow.

In this paper, we are interested in obtaining the phenomenological
important contributions in hadron production in
the forward region of the polarized beam.
Similar to the twist-three distribution function
contributions, we find that the twist-three fragmentation function
contributions also contain a derivative term which certainly
will dominate this part of contribution in the
forward rapidity region. In addition, the derivative contribution
from the twist-three fragmentation function is associated with
$\hat H(z)$ defined in Eq.~(\ref{hzdef}), whereas
the contribution associated with $\hat{H}_F(z_1, z_2)$ does
not contain any derivative contributions. In this paper, we
will report the derivative-term results, and carry out the numerical estimates
for their contributions to the single spin asymmetries
at RHIC. We will leave the detailed derivations, including
the derivative and non-derivative terms in a future publication.

\begin{figure}[t]
\begin{center}
\includegraphics[height=4.0cm,angle=0]{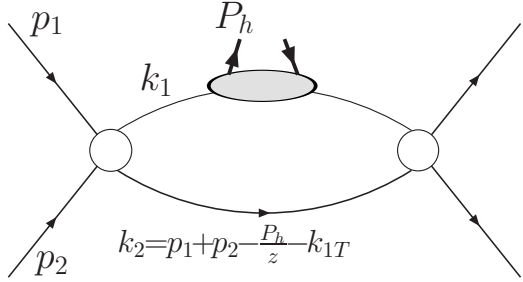}
\vspace*{-0.2cm} \epsfysize=4.2in
\end{center}
\caption{\it Generic Feynman diagrams to calculate the derivative term
of the twist-three fragmentation function contributions to the single inclusive
hadron production in $pp$ scattering, $p^\uparrow p\to hX$:
$p_1$ and $p_2$ are the two incident partons' momenta,
$k_1$ and $k_2$ are the outgoing partons' momenta, where the quark $k_1$
fragments into final-state hadron $P_h$. The expansion of the
scattering amplitude in terms of transverse
momentum component of $k_1=P_h/z+k_{1T}$ leads to the twist-three
contributions from the fragmentation function
$\hat H(z)$ defined in Eq.~(\ref{hzdef}).
}
\label{illustrate}
\end{figure}

We follow the technique developed in the last few years
\cite{qiu, Kouvaris:2006zy,{Koike:2009ge}, Kang:2008qh, Eguchi:2006qz, Ji:2006ub, yuan-zhou}
to derive the contributions from the twist-three fragmentation functions.
First, we calculate the associated scattering amplitudes in terms of
various twist-three matrix elements of hadrons. From diagrammatic
point of view, these individual contributions are not in a gauge invariant
way. However, the final results will be gauge invariant when we sum up
all the contributions. In terms of twist-three operators,
we have the contributions from the $\partial_T \psi$, $A_T$,
and $\partial_T A^+$ ($\partial ^+A_T$). These contributions
indeed will form the gauge invariant results in terms of $\hat H(z)$,
$\hat H_F$ as defined above. In particular,
$\partial_T \psi$ term corresponds to $\hat H(z)$ whereas
$A_T$ term corresponds to $\hat H_F$.
Since the derivative terms are associated with the function
$\hat H(z)$, we will need to calculate the contributions from
the $\partial_T\psi$ part. In order to carry out this
calculation, we have to perform the collinear expansion
for the hard partonic part associated with this matrix element.
In Fig.~\ref{illustrate}, we illustrate the generic Feynman diagrams contributing
to the derivative terms.
In this diagram, $p_1$ and $p_2$ are the two incoming partons' momenta
from the polarized and unpolarized nucleons, respectively;
$k_1$ and $k_2$ are the two outgoing partons' momenta where $k_1$
fragments into the final-state hadron $P_h$.
The derivative terms come from the transverse momentum
expansion of hard partonic scattering amplitudes, in particular,
from the on-shell condition for the unobserved particle $k_2$ in Fig.~\ref{illustrate}.
This momentum can be written as $k_2=p_1+p_2-k_1$, where
we can parameterize $k_1$ as $k_1=P_h/z+k_{1T}$. Thus, the
on-shell condition will lead to the following expansion,
\begin{equation}
\delta((p_1+p_2-k_1)^2)=\delta((p_1+p_2-P_h/z)^2)-2(p_1+p_2)\cdot k_{1T}\delta'((p_1+p_2-P_h/z)^2) \ ,
\end{equation}
where the second term will result into a derivative term contribution to the
transverse spin dependent differential cross section.
As discussed above, for convenience, we use a light-like
vector $n_h$ to constrain the transverse momentum expansion
which is perpendicular to the final state hadron's momentum.
Therefore, $k_{1T}$ can be expressed as
$k_{1T}^\mu=k_1-k_1\cdot P_h n_h^\mu-k_1\cdot n_h P_h^\mu$.
Substituting this expression into the above equation,
we will obtain the contribution from the matrix element
associated with the operator $\partial_T\psi$ which will
result into the contribution of $\hat H(z)$.
The derivative on the Delta function will be translated
into the derivative on the twist-three fragmentation
function $\hat H(z)$ by partial  integral.
This contribution is similar to what has been calculated for the twist-three
distribution functions~\cite{qiu}. In particular, the hard part is proportional to
that from the relevant twist-two Born diagrams. In our case,
it is  the transverse spin transfer coefficients calculated
in Ref.~\cite{werner-marco}. However, there exist major
difference between these two contributions. The twist-three
distribution contributions come from the pole contributions
from the initial/final state interactions, and the derivative
terms have contributions from the expansion of the on-shell
condition for the unobserved particle and the double pole
contribution from the final state interaction diagrams~\cite{qiu}.
In the twist-three fragmentation function contributions,
in contrast, we do not take pole contributions, and the derivative
terms only come from the expansion of the on-shell
condition of the unobserved particle in the above $2\to 2$
processes. The final result for the derivative contributions of $\hat H(z)$
takes the following form,
\begin{eqnarray}
E_h \frac{d^3 \Delta \sigma(s_\perp)}{d^3P_h} &=&
\epsilon_{\perp \alpha \beta} S_\perp^\alpha
\frac{ 2 \alpha_s^2}{S}
\sum_{a,b,c}
\int^1_{x'_{min}} \frac{dx'}{x'}f_b(x') \frac{1}{x} h_a(x)\int^1_{z_{min}} \frac{dz}{z}
\left [ -z \frac{\partial}{\partial z}\left(\frac{\hat{H}(z)}{z^2}\right) \right ]\nonumber \\
&&
\times
\frac{1}{x'S+T/z}
\left\{\frac{P_h^\beta+z(p_2\cdot P_hn_h^\beta-p_2\cdot n_hP_h^\beta)}{-z\hat u}\right\}
H_{ab \rightarrow c}( \hat{s}, \hat{t}, \hat{u} )\ ,
\end{eqnarray}
where
\begin{equation}
x'_{\rm min}= \frac{-T/z}{S+U/z} \  ,  ~~~~~~z_{\rm min}=
-\frac{T+U}{S} \ ,
\end{equation}
and the hard factors $H_{ab\to c}$ are defined as
\begin{eqnarray}
H_{qq'\to qq'}&=&H_{q\bar q'\to q\bar q'}=
\frac{N_c^2-1}{4N_c^2}\frac{4\hat s\hat u}{-\hat t^2} \ ,
\nonumber\\
H_{qq\to qq}&=&
\frac{N_c^2-1}{4N_c^2}\left[\frac{4\hat s\hat u}{-\hat
t^2}-\frac{1}{N_c}\frac{4\hat s}{-\hat t}\right] \ ,
\nonumber\\
H_{q\bar{q}\to q\bar{q}}&=&H_{\bar{q}q\to \bar{q}q}=
\frac{N_c^2-1}{4N_c^2}\left[\frac{4\hat s\hat u}{-\hat
t^2}+\frac{1}{N_c}\frac{4\hat u}{\hat t}\right] \ ,
\nonumber\\
H_{\bar{q}q\to q\bar{q}}&=&H_{q\bar{q}\to \bar{q}q}
=-\frac{N_c^2-1}{N_c^3}\ ,
\nonumber\\
H_{qg\to qg}&=&
\frac{N_c^2-1}{N_c^2}+\frac{1}{2}\frac{4\hat s\hat
u}{-\hat t^2} \ ,
\end{eqnarray}
where $\hat s$, $\hat t$, and $\hat u$ are the usual partonic
Mandelstam variables. They are the same as the transverse
spin transfer hard coefficients calculated in Ref.~\cite{werner-marco}.
Substituting the expression of $n_h$ into Eq.~(8), the factor
in the bracket can also be written as,
\begin{equation}
\left\{\frac{P_h^\beta+z(p_2\cdot P_h
n_h^\beta-p_2\cdot n_hP_h^\beta)}{-z\hat u}\right\}\to
\left(\frac{P_h^\beta}{z}\right)\frac{x-x'}{x(-\hat u)+x'(-\hat t)} \ .
\end{equation}
In the forward rapidity region of the polarized nucleon,
we have $x\gg x'$ and $-\hat u\gg -\hat t$,
and we can further simplify the transverse spin dependent
differential cross section as
\begin{eqnarray}
E_h \frac{d^3 \Delta \sigma(s_\perp)}{d^3P_h}|_{forward}&=&
\epsilon_{\perp \alpha \beta} S_\perp^\alpha P_{h\perp}^\beta
\frac{ 2 \alpha_s^2}{S}
\sum_{a,b,c}
\int^1_{x'_{min}} \frac{dx'}{x'}f_b(x') \frac{1}{x} h_a(x)\int^1_{z_{min}} \frac{dz}{z}
\left [ -z \frac{\partial}{\partial z}\left(\frac{\hat{H}(z)}{z^2}\right) \right ]\nonumber \\
&&\times
\frac{1}{x'S+T/z}\frac{1}{-z\hat u}
H_{ab \rightarrow c}( \hat{s}, \hat{t}, \hat{u} )\ .
\end{eqnarray}
This term will be most phenomenological relevant for the single
spin asymmetries of hadron production in the forward
direction of the polarized nucleon.

In order to demonstrate the twist-three fragmentation function contribution
to the SSA in inclusive hadron
production in $p^\uparrow p$ collisions, we need the unknown, but universal, twist-three
fragmentation function $\hat{H}(z)$. We notice that $\hat{H}(z)$ can be related to
the Collins fragmentation function $H_1^\perp(z, p_\perp^2)$ as in Eq.~(\ref{hz}), which
has been studied from the available experimental data~\cite{Vogelsang:2005cs, Anselmino:2007fs}.
However, we emphasize that the Collins function $H_1^\perp(z, p_\perp^2)$ are 
fitted from small transverse momentum region where TMD factorization applies. 
To obtain the functional form for $\hat H(z)$, one has to assume a transverse
momentum dependence in the Collins fragmentation functions. 
In principle, the twist-three fragmentation functions $\hat{H}(z)$ should be 
extracted from the experimental data of the SSAs at large
transverse momentum region where collinear factorization applies, similar to what has
been done in~\cite{Kouvaris:2006zy}, or from the transverse momentum weighted azimuthal
asymmetry measurements in which the twist-three fragmentation function $\hat H(z)$
enters directly.

For the purpose of estimating its contribution to the SSAs in 
the inclusive $\pi^0$ production in $p^\uparrow p$ collisions 
at RHIC energy, we parameterize the twist-three
fragmentation function in terms of the leading-twist fragmentation
function,
\begin{equation}
\hat H(z)=C_f z^aD(z)\ .
\label{parHz}
\end{equation}
We leave a more comprehensive parameterization for these fragmentation
functions in a future study, including that for charged mesons as well.
The additional $z$ factor in the parameterization comes from the
consideration that this novel fragmentation is mostly a valence
behavior. Normally, we would also have a $(1-z)$ suppression for the
twist-three function. However, for scalar hadron production,
the twist-three fragmentation function is not power suppressed
in terms of $(1-z)$, similar to the power counting of the
Boer-Mulders function of $\pi$ meson at large-$x$~\cite{Brodsky:2006hj}.
We have also adopted the quark transversity distributions
from the parameterizations in Ref.~\cite{Martin:1997rz} and the unpolarized
fragmentation function in \cite{Kretzer:2000yf}.
In Fig.~\ref{anplot}, we show the predictions with the above parameterizations
for the $\pi^0$ production with $C_f=-0.4$. The three curves from up to bottom
correspond to $a=1$ (solid), $a=2$ (dashed), and $a=4$ (dotted), respectively.
With our parametrization of $\hat{H}(z)$, the twist-three fragmentation function
can generate a sizable SSA in inclusive $\pi^0$ production
at RHIC energy $\sqrt{s}=200$ GeV. These contribution are comparable to that of the
twist-three distribution from the polarized nucleon which have
similar parameterizations~\cite{qiu,Kouvaris:2006zy}.
\bef
\psfig{file=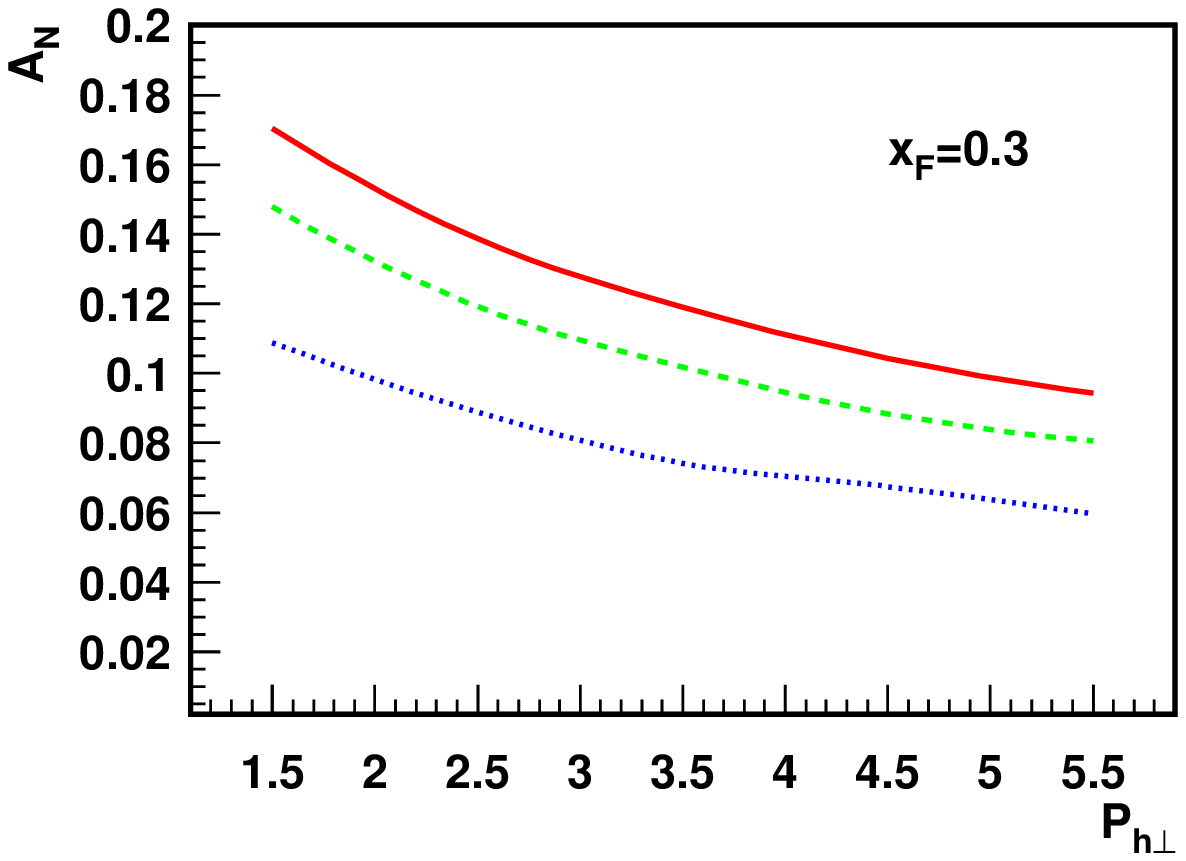, width=2.9in}
\hskip 0.2in
\psfig{file=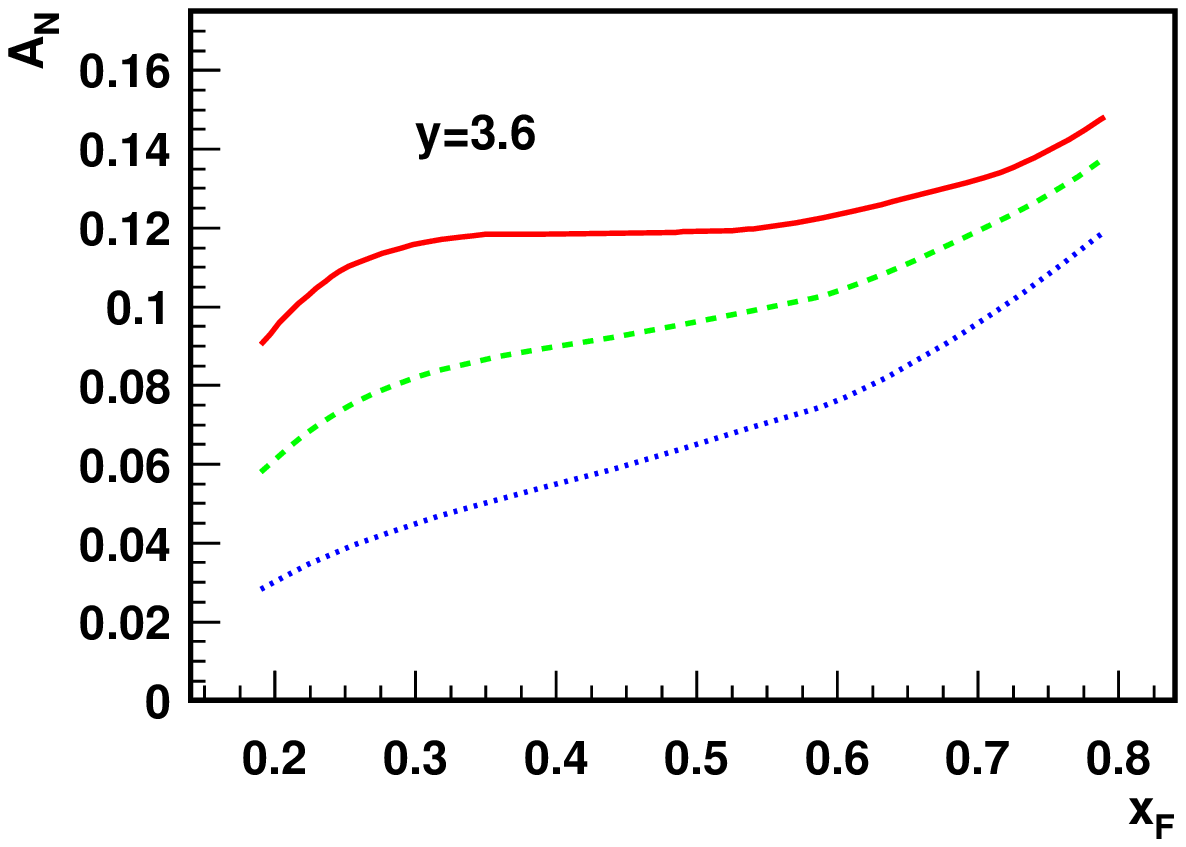,width=2.9in}
\caption{\it Twist-three fragmentation contributions to the single spin
asymmetries in hadron production in the forward direction of the polarized
nucleon at RHIC at $\sqrt{s}=200GeV$, as functions of $x_F$,
with the parameterization for the twist-three fragmentation function
$\hat H(z)$ in Eq.~(\ref{parHz}) with parameters $C_f=-0.4$ GeV and $a=1,2,4$
(from up to bottom), respectively. }
\label{anplot}
\eef

The single transverse spin asymmetry of $\eta$ meson has also been studied by
the STAR collaboration at RHIC recently \cite{steve}.
A significantly larger asymmetry $A_N$ has been observed
 for $\eta$ meson compared to $\pi^0$.
As we discussed, in the twist-three collinear factorization, the dominate
contribution comes from the twist-three distribution of the
polarized nucleon, as shown in the first term of Eq.~(\ref{collinear}), 
and the twist-three fragmentation function, as shown in the third term 
of Eq.~(\ref{collinear}). The difference
between $\eta$ and $\pi^0$ from the first term will be mainly due to the difference in the
unpolarized meson fragmentation function. However, from
the measurements of spin-averaged cross section of $\eta$ and $\pi^0$ by PHENIX
collaboration at RHIC \cite{Adler:2006bv}, one finds that for a large range of transverse
momentum, $\eta/\pi^0$ ratio is a constant. This indicates that the unpolarized
fragmentation function for $\eta$ and $\pi^0$ is very similar, and will lead to the similar
SSAs for them if this contribution dominates.
In other words, it will be very difficult to explain the large difference
between the SSAs of $\eta$ and $\pi^0$ mesons from the twist-three 
distribution contribution from the polarized nucleon\footnote{
It might still be possible that the strange quark contribution
from the polarized nucleon may dominates and leads to a larger SSA in
$\eta$ meson production, which is, however, unlikely in the forward
rapidity region (the valence region) of the polarized nucleon.}.
Since the second term in Eq.~(\ref{collinear}) generates a very small asymmetry,
one would expect the large difference between $\eta$ and $\pi^0$ would
come from the twist-three fragmentation function contribution
if one believes the collinear factorization is indeed the right approach to describe the SSA
of inclusive hadron production in $pp$ collisions. Since the twist-three fragmentation
function $\hat{H}(z)$ for $\eta$ and $\pi^0$ in general need not to be the same, this
might generate the needed difference observed by the experiments if $\eta$ meson
has a much larger twist-three fragmentation function $\hat{H}(z)$ compared to $\pi^0$.
To test this scenario, it will be very important to study the associated Collins
fragmentation for $\eta$ meson in $e^+e^-$ annihilation and/or semi-inclusive DIS
processes and compare to that for $\pi^0$ meson. We hope that, in particular, the
BELLE collaboration can carry out this measurement and cross check with the
STAR observation.
Meanwhile, we emphasize that to better understand the single spin asymmetry
for these processes and finally pin down the difference, one needs to
take into account both twist-three contributions and perform a global analysis
of the experimental data.

In conclusion, in this paper, we have studied the twist-three fragmentation function
contribution to the inclusive hadron's SSA in $pp$ scattering $p^\uparrow p\to hX$. With
a simple parametrization for the twist-three fragmentation function, we estimate its
contribution to the SSAs of $\pi^0$ production at RHIC energy. We find that the
contribution of the twist-three fragmentation function is comparable to that of the
twist-three distribution function from the polarized nucleon. We comment on the
possibility to use our approach to describe the large difference of the SSAs between
the $\eta$ and $\pi^0$ meson. We emphasize that one need to include both
contributions from twist-three distribution and fragmentation functions
into a global analysis, in order to better understand the single spin
asymmetry for the inclusive hadron production.

We thanks Jianwei Qiu and Yuji Koike for helpful discussions.
This work was supported in part by the U.S. Department of Energy
under contract DE-AC02-05CH11231. We are grateful to RIKEN,
Brookhaven National Laboratory and the U.S. Department of Energy
(contract number DE-AC02-98CH10886) for providing the facilities
essential for the completion of this work.

\end{document}